\documentclass[twocolumn,preprintnumbers,amsmath,amssymb,floatfix,superscriptaddress,nofootinbib]{revtex4}
\usepackage{amsmath,hyperref,amssymb}
\usepackage{graphicx}
\usepackage{dcolumn}
\usepackage{bm}
\usepackage{verbatim}
\usepackage{tabularx}
\usepackage{slashed}
\usepackage{cancel}
\usepackage{hyperref}

\def\be{\begin{equation}}
\def\ee{\end{equation}}
\def\bea{\begin{eqnarray}}
\def\eea{\end{eqnarray}}
\def\ba#1\ea{\begin{align}#1\end{align}}
\def\bg#1\eg{\begin{gather}#1\end{gather}}
\def\bm#1\em{\begin{multline}#1\end{multline}}
\def\bmd#1\emd{\begin{multlined}#1\end{multlined}}

\newcommand{\mc}{\mathcal}
\renewcommand{\t}{\tilde}

\begin{document}

\title{The  a-theorem and the Markov property of the CFT vacuum}
\author{Horacio Casini, Eduardo Test\'e, Gonzalo Torroba}
\affiliation{
   Centro At\'omico Bariloche and CONICET,\\
   S.C. de Bariloche, R\'io Negro, R8402AGP, Argentina
   }

\date{\today}
\begin{abstract}
We use strong sub-additivity of entanglement entropy, Lorentz invariance, and the Markov property of the vacuum state of a conformal field theory to give a new proof of the irreversibility of the renormalization group in $d=4$ space-time dimensions -- the $a$-theorem. This extends the proofs of the $c$ and $F$ theorems in dimensions $d=2$ and $d=3$ based on vacuum entanglement entropy, and gives a unified picture of all known irreversibility theorems in relativistic quantum field theory.
\end{abstract}

\maketitle

\section{Introduction}\label{sec:intro}

The central idea of the renormalization group is that the change of physics with scale in a quantum field theory (QFT) can be assimilated to a change in the parameters of the Hamiltonian describing the relevant degrees of freedom. This flow in the space of theories brings us from the ultraviolet (UV) fixed point at short scales to the infrared (IR) one at large scales. At the fixed points, the physics stops changing, and we focus on relativistic systems in $d$ spacetime dimensions, where the endpoints of the flow are conformal field theories (CFT).

It has long been known that the renormalization group (RG) is irreversible in two spacetime dimensions \cite{Zamolodchikov:1986gt}. This result, known as the $c$-theorem, shows that the conformal anomaly $c$ (a dimensionless quantity depending on the CFT) decreases between the UV and IR fixed points. The value of $c$ at conformal fixed points is thus interpreted as a precise measure of the number of field degrees of freedom; Zamolodchikov's theorem then realizes the intuitive idea that this number should decrease at larger scales due to the decoupling of massive modes. It also establishes an ordering of CFTs: theories with smaller $c$ in the UV cannot flow to theories with larger $c$ in the IR, and the renormalization group is irreversible. 

In four spacetime dimensions, Cardy \cite{Cardy:1988cwa} gave arguments suggesting that a particular coefficient of the conformal anomaly, the $a$ coefficient of the Euler term, should also decrease under the RG.  After long being sought, the $a$-theorem was proved by \cite{Komargodski:2011vj}. 

For odd dimensions the situation was initially unclear because there are no conformal anomalies. Based on RG irreversible quantities in holography, Ref. \cite{Myers:2010xs} proposed that in odd dimensions the relevant monotonous quantity is the constant term of the entanglement entropy of a sphere. This conjecture, now known as the $F$-theorem, was established for $d=3$ in \cite{Casini:2012ei}, extending the proof \cite{Casini:2004bw} of the $c$-theorem in $d=2$. The crucial property here is the strong sub-additivity of entropy, which ultimately gives a different perspective on unitarity and irreversibility. In a related development in supersymmetric QFTs, \cite{Jafferis:2011zi} conjectured that the constant term in the free energy of a 3-sphere is monotonous -- hence the name $F$. In fact, this quantity is the same as the constant term of the entanglement entropy of a sphere \cite{Casini:2011kv}, and the proposals of \cite{Myers:2010xs} and \cite{Jafferis:2011zi} actually coincide.   

These developments suggest that in any dimension the monotonous quantity is the universal part of the entanglement entropy of a sphere. This is proportional to the Euler anomaly for even dimensions. While this points to some underlying principle behind the irreversibility of the RG across dimensions (see e.g. \cite{Giombi:2014xxa}), so far the techniques employed have been quite specific to each particular dimension.
Only an entropic proof exists for $d=3$, and so far only a proof based on local field theoretic quantities was known in $d=4$; both entropic and correlator techniques can be used to prove the theorem in $d=2$. An important difficulty for proofs based on correlations functions in odd dimensions is that the $F$ quantity is, in contrast to anomalies, a rather nonlocal object.    

In this work we prove the $a$-theorem using entropic techniques, and provide a unifying approach to the irreversibility of the RG. The new key ingredient here will be the recently discovered Markovian property of the vacuum state of a CFT \cite{Casini:2017roe}. Based on this we will extend the approach in \cite{Casini:2012ei} to $d=4$, resolving previous obstacles from problematic terms in the entanglement entropy (EE) of unions and intersections of spheres.

\section{The setup}

We consider a RG flow between UV and IR CFT fixed points in $d$ spacetime dimensions. The flow is triggered by a perturbation with some relevant operator $\mc O$ of dimension $\Delta<d$,
\be\label{eq:S1}
S_1= S_0+\int d^dx\,g \, \mathcal O(x)\,.
\ee
The theory at the UV fixed point is denoted by $\mc T_0$, while $\mc T_1$ is the theory (\ref{eq:S1}).
In order to understand the irreversibility of the RG, we will study the entanglement entropy on spheres.
Let $\rho_X$ be the reduction of the global state to the region $X$ and $S(X) = -\text{Tr} (\rho_X \log \rho_X)$  its von Neumann entropy. This is the entanglement entropy between $X$ and the complementary region $\bar{X}$, which we seek to compute.
 
For the vacuum state of a QFT, the EE of a sphere is in general a complicated function of the radius $r$, a distance cut-off $\epsilon$, and the dimensionful parameters of the theory. At fixed points and for a sufficiently geometric cut-off (such as \cite{Casini:2015woa}) the entropy simplifies to 
\bea
S(r)&=&\mu_{d-2}\,r^{d-2}+\mu_{d-4}\, r^{d-4}+\cdots \nonumber 
\\ &&\hspace{1cm}+ \left\lbrace \begin{array}{ll} (-)^{\frac{d}{2}-1} 4\,A\, \log(R/\epsilon)\,& d\;  \textrm{even}\,.\\ (-)^{\frac{d-1}{2}} F\,& d\,\,\textrm{odd} \,. \end{array}\right.\label{even}
\eea    
See e.g. \cite{Metlitski:2011pr, Liu:2012eea, Liu:2013una}.
The last term gives the universal part of the EE. $A$ is the Euler trace anomaly coefficient for even dimensions \cite{Solodukhin:2008dh}, and $F$ is the constant term of the free energy of a $d$-dimensional Euclidean sphere. 

The reason for this expression is that the large distance entanglement does not change with dilatations at a fixed point (with the exception of the anomaly term), and hence the $r$-dependence comes from contributions that are local on the entangling surface, i.e. integrals of curvature tensors. Curvature tensors with odd number of dimensions change sign when they are evaluated on the two sides of the entangling surface and cannot appear in the expansion because of the identity of entropies for complementary regions $S(X)=S(\bar{X})$. Hence, only powers below the area term differing by an even number appear in (\ref{even}).

The coefficients $\mu_{d-k}$ have dimension $d-k$.
For a CFT (such as $\mc T_0$ above), the only dimensionful parameter is the cutoff $\epsilon$, so that $\mu_{d-k}\sim \epsilon^{-(d-k)}$. For the theory $\mc T_1$ with the relevant perturbation (\ref{eq:S1}) the situation is richer. For small spheres $r \sim 0$, where we can apply conformal perturbation theory near the UV, we expect 
\be\label{eq:mupert}
\mu_{d-k}^{UV} \sim \epsilon^{-(d-k)}+g^2 \epsilon^{-(d-k)+2(d-\Delta)}+\ldots
\ee 
This is UV divergent (and perturbatively computable) for $\Delta \ge (d+k)/2$. Additionally, for small $r$ we expect finite perturbative corrections to the entropy of the form $S(r) \sim g^2 r^{2(d-\Delta)}$, which are nonlocal. See \cite{Liu:2012eea} for holographic examples. On the other hand, taking $r\to \infty$ the IR fixed point is approached; besides the UV divergent terms just discussed, the EE will contain finite renormalizations to $\mu_{d-k}^{IR}$. These contributions, which should be regularization independent, depend on the full RG flow, and are generally nonperturbative. Nonlocal corrections, however, are absent at the IR fixed point.

\section{Irreversibility from strong sub-additivity}   

\begin{figure}[t]
\begin{center}
\includegraphics[width=0.35\textwidth]{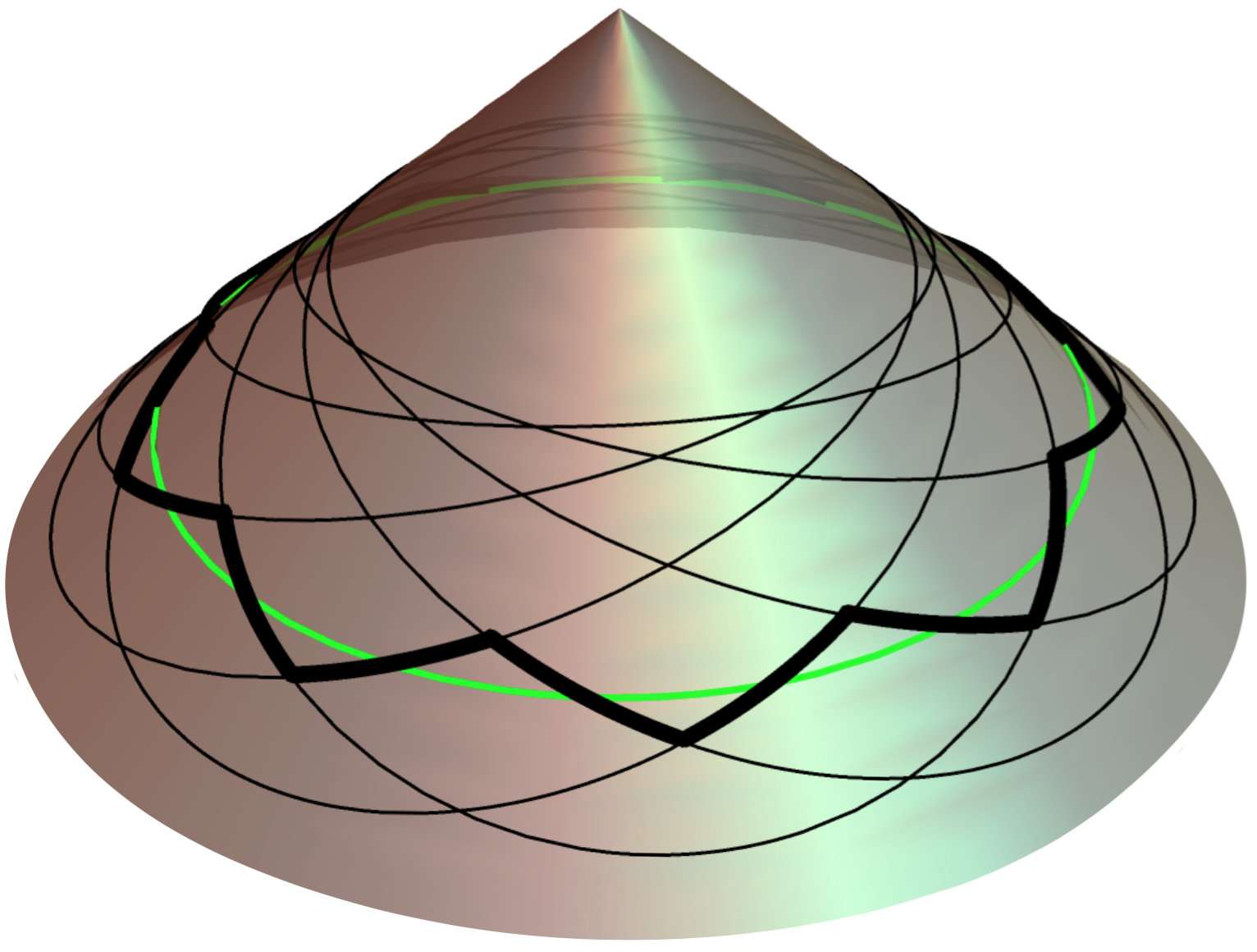} 
\caption{Boosted and uniformly distributed circles lying on the null cone in $d=3$ spacetime dimensions. The vertical axis of the cone gives the time direction. A wiggly sphere, corresponding to one of the sets in (\ref{ecu}), is highlighted in black, and its corresponding limiting circle is highlighted in green.}
\label{conosolo}
\end{center}
\end{figure}

The idea is to relate EE coefficients of the UV and IR fixed points using a property of entropy called 
the strong subadditivity inequality (SSA) \cite{lieb1973proof}. For two regions $A$ and $B$ it reads
\begin{equation}\label{SSA} 
S(A)+ S(B)\ge S(A\cap B)+ S(A \cup B)\ .
\end{equation}
This motivates the construction in \cite{Casini:2012ei} of the geometrical setup illustrated in Fig. \ref{conosolo}. A large number of rotated copies $X_i$, $i=1,\cdots,N$ of a boosted sphere  are placed on a null cone. All these spheres are chosen to have the same radius $\sqrt{R r}$, and are equally distributed in the angular directions. 
The $t=0$ projection of these spheres lies between radii $r$ and $R$.
Repeated use of the SSA gives
\begin{eqnarray}\label{ecu}
\sum_{i}S(X_{i}) &\geq& S(\cup _{i}X_{i})+S(\cup _{\{ij\}}(X_{i}\cap
X_{j})) \\
&&+S(\cup _{\{ijk\}}(X_{i}\cap X_{j}\cap X_{k}))+...   +S(\cap _{i}X_{i}) \,. \nonumber 
\end{eqnarray}
There are $N$ terms on each side of (\ref{ecu}). The right hand side contains entropies of regions that approach spheres for large $N$ but have wiggly boundaries in a null direction. The aim is to get inequalities involving only spheres in the limit. 

The main question is then how to relate entropies of wiggly spheres with those of regular spheres. Since the surfaces are on the lightcone, the area term along the boundary of a wiggly sphere matches that of a regular sphere passing through the middle of the wiggles; see Fig. \ref{conosolo}.  However, the local curvature is different, and so generically we do not expect the entropies to agree (we will see an example below). Unfortunately, a direct calculation of the wiggly contributions seems too complicated, and a different route is needed. It is important to realize, however, that the differences in the EE of wiggly and regular spheres are purely UV at large $N$. If we managed to subtract the UV contributions while still maintaining strong sub-additivity, the wiggly contributions would go smoothly to regular contributions. This is the point where the recently discovered Markov property \cite{Casini:2017roe} comes into play.

For any two regions $A$ and $B$ with boundary lying on the lightcone, the CFT vacuum in any dimension is a Markov state, namely, it saturates the SSA inequality \cite{Casini:2017roe}
\be
S(A)+S(B)-S(A\cap B)-S(A\cup B)=0\,. \label{sur}
\ee
This follows from the form of the modular Hamiltonian on the lightcone, as well as from algebraic QFT methods, but at first it looks rather surprising. Indeed, intersections and unions of regions contain additional local singularities that may produce divergent terms in the entropy, see e.g. \cite{Klebanov:2012yf, Myers:2012vs}.

Let us then briefly describe how this works out in $d=4$, where all interesting features already appear. The area term always cancels in the combination (\ref{sur}), as is the case for the $\log(\epsilon)$ term coming from a local integral of the curvature  outside the singular points from intersections. This would also hold for spheres in a plane. A new feature comes from the intersection of two spheres; it gives a term that scales with the length $\ell$ of the line of intersection as $\ell/\epsilon$. This must be an integral along this line that is locally the same as the one of the intersection of two planes tangent to the spheres at a point of the intersection. These two spatial planes are contained in a null hyperplane of dimension $3$. Hence we can boost one of the planes with a boost that keeps the other plane fixed and the null hyperplane invariant. There is then no local notion of angle between the two planes -- this feature cannot contribute since we have no local geometrical quantity to distinguish it from two parallel planes. Next, the intersection lines are curved and can produce a $\log(\epsilon)$ contribution times a line integral of the curvature. This cannot be eliminated by boosting but we note that it is a signed curvature; the union and intersection of two spheres have exactly opposite contributions of this form and hence cancel out. Finally, we have the vertices where three spheres intersect. This trihedral angle is immersed in a null hyperplane, and does not contribute by the same boost argument as before.

Because of the Markov property, the difference in EE between the CFT $\mc T_0$ and the theory $\mc T_1$ along the flow,
\be
\Delta S(r) = S_{\rho^1}(r)-S_{\rho^0}(r)
\ee
still satisfies the strong sub-additivity (\ref{SSA}), and (\ref{ecu}) applies to $\Delta S$. In this way, all UV effects associated to wiggles cancel out from the inequality (recall that we take $N \to \infty$ with fixed coupling $g$) and $\Delta S_{wiggly}$ can be replaced by $\Delta S_{regular}$ inside the SSA formula.

The wiggly spheres lie approximately on constant $t$ planes, with radius $l$ ranging from $r$ to $R$. Let $l_k$ be the radius of the wiggly sphere of order $k$, that is, the one formed by the union of the intersections of $k$ spheres. Defining the density of wiggly spheres 
\be
\beta(l)=\frac{1}{N}\frac{dk}{dl}\,,
\ee
 the geometry gives \cite{Casini:2012ei}
\begin{equation}
\beta(l)=  \frac{2^{d-3}\Gamma[(d-1)/2]}{\sqrt{\pi}\Gamma[(d-2)/2]}\,  \frac{(r R)^{\frac{d-2}{2}} \left( (l-r)(R-l) \right)^{\frac{d-4}{2}}    }{ l^{d-2}  (R-r)^{d-3}}\ .
\end{equation}
Hence the inequality becomes 
\be
\Delta S(\sqrt{r R})\ge \frac{1}{N}\sum_{k=1}^N \Delta S_k\approx \int_r^R dl\, \beta(l)\, \Delta S(l)\,, \label{wig}
\ee
where at large $N$ the sum is replaced by an integral, and we have already replaced the contribution $\Delta S$ from wiggly spheres by that of regular spheres. Finally, expanding for small $R-r$ we arrive to our main result,
\be\label{eq:main}
r  \,\Delta S''(r) - (d-3)\, \Delta S'(r) \leqslant 0\ .
\ee

\section{The entropic A-theorem}

Before proving the $a$-theorem, let us discuss the implications of this inequality in lower dimensions.

For $d=2$ (\ref{eq:main}) gives
\be
 (r\, \Delta S'(r))'\leqslant 0\,.\label{dec}
\ee
In fact, this is valid directly for $S(r)$ since wiggly spheres are just ordinary intervals. 
Defining $\Delta c(r)=c(r)-c_{UV}=r \,\Delta S'(r)$, this gets the coefficient of the logarithmic term in the entropy for fixed points. Since it decreases with size, (\ref{dec}) gives a proof of  the $c$-theorem. 

For $d=3$, (\ref{eq:main}) becomes $(\Delta S(r))''\leqslant 0$ and this has two implications. First, it gives an ``area theorem'', implying that the quantity 
\be
\Delta \t \mu_1(r) \equiv \Delta S'(r)
\ee
decreases along the flow. This is finite for $\Delta<5/2$, and coincides with the subtracted area coefficient at fixed points. Hence $\Delta \mu_1^{IR}\le \Delta \mu_1^{UV}$. For larger $\Delta$, the nonlocal UV term discussed below (\ref{eq:mupert}) dominates, making $\Delta \t \mu_1$ diverge as $r\to 0$.\footnote{The area theorem in $d$ dimensions was proved using positivity of the relative entropy in \cite{Casini:2016udt}.} The other consequence of the inequality is that
\be
 (r \,\Delta S'(r)-\Delta S(r))'\leqslant 0\,.
\ee    
The CFT contribution drops out (both the area and constant term cancel out), and hence the quantity $F(r)=r S'(r)-S(r)$ decreases monotonically and agrees with $F$ at fixed points. This gives a proof of the $F$-theorem; it agrees with that in \cite{Casini:2012ei}, where the wiggly circles were replaced by regular ones because in $d=3$ the wiggles do not contribute to the SSA inequality.

Finally, let us consider $d=4$. The CFT contribution is
\be
S_{\rho^0}(r)=\mu^0_2\, r^2-4 A_{UV} \log(r/\epsilon)\,,\label{4d}
\ee
where $A_{UV}$ is the $a$-anomaly coefficient of the UV fixed point. Replacing this into (\ref{eq:main}) obtains
\be\label{eq:main2}
r S_{\rho^1}''(r)- S_{\rho^1}'(r) \le \frac{8 A_{UV}}{r}\,.
\ee
Evaluating the left hand side at the IR fixed point gives
\be
A_{IR}\leqslant A_{UV}\,.
\ee
This completes our proof of the $a$-theorem using entropic techniques.

Let us emphasize that this is the point where the Markov property of the CFT plays a key role. Had we just replaced wiggly contributions by regular contributions to the entropy (instead of doing it for $ \Delta S$), we would have obtained that the left hand side in (\ref{eq:main2}) is nonpositive. And this is violated at fixed points. Therefore, we see explicitly in this case that the entropy contributions of wiggly spheres do not tend smoothly to those of regular spheres. With our present approach we have avoided this problem by using the strong sub-additivity property of $\Delta S$. Therefore the Markov property of the CFT vacuum is essential for obtaining the $a$-theorem.

Let us end with two remarks. First, an analog to a c-function can be written as $\Delta c(r)=r \,\Delta S'(r)-2 \Delta S(r)$. It is decreasing, it vanishes at the UV, and at the IR it approaches
\be
\Delta c\approx 8 (A_{IR}-A_{UV}) \log(Mr)\,, 
\ee
where $M$ is some scale of the RG. It does show the decrease of $A$; however, it does not converge to a finite value for large $r$. Finally, as for $d=3$ we have here also an area theorem. Defining the quantity
\be
\Delta \t \mu_2(r)= \frac{\Delta S'(r)}{2r}=\frac{1}{2r}\left(S_{\rho^1}'(r)-S_{\rho^0}'(r) \right)\,,
\ee
this is always decreasing $\Delta \t \mu_2'(r)\leqslant 0$. For $\Delta<3$ it is finite and approaches
the subtracted area coefficient at fixed points. Hence $\Delta \mu_2^{IR} \leqslant \Delta \mu_2^{UV}$. In $d=2$ the area theorem coincides with the $c$-theorem, as discussed in \cite{Casini:2016udt}.

\section{Extension to higher dimensions and final remarks}

For dimensions higher than $4$ we have more than two coefficients of the entropies $S_{\rho^0}$ and $S_{\rho^1}$ in the IR. Eq. (\ref{eq:main}) gives two relevant inequalities. The first is for the area term. This follows from the interpolating quantity 
\be
\Delta \t \mu_{d-2}(r)=\frac{\Delta S'(r)}{(d-2)r^{d-3}}
\ee
that always decreases. From (\ref{even}), the structure of UV divergences ignoring order one coefficients is
\be
\Delta \t \mu_{d-2}(r)= g^2 \epsilon^{d+2-2\Delta}\left(1+ \frac{\epsilon^2}{r^2}+\ldots \right)+\text{finite}\,.
\ee
In the UV $r \ll g^{-1/(d-\Delta)}$, the finite term is of order $g^2 r^{d+2-2\Delta}$. Near the IR fixed point we expect, on dimensional grounds, a finite term of order $g^{(d-2)/(d-\Delta)}$.

Therefore, $\Delta \t \mu_{d-2}(r)$
is finite for $\Delta<(d+2)/2$ and interpolates between area terms, so $\Delta \mu_{d-2}^{IR} \leqslant \Delta \mu_{d-2}^{UV}$. However, if 
$\Delta\ge(d+2)/2$, $\Delta \tilde{\mu}_{d-2}(r)$ is divergent, while its change with $r$ can still be finite if $\Delta < (d+4)/2$. The total running of this quantity from $r=0$ to $r=\infty$ is infinite due to the finite terms in the UV.

The other inequality comes from observing the IR value of (\ref{eq:main}). This is dominated by the next leading term proportional to $r^{d-4}$ in the entropies and gives 
\be\label{eq:result2}
\Delta \mu_{d-4}^{IR}\geqslant 0\,.
\ee
This is finite or not according to whether $\Delta < (d+4)/2$ or $\Delta\ge (d+4)/2$ respectively. For $d=4$ this gives the $a$-theorem discussed before.

The area term is related to the renormalization of Newton's constant. Along similar lines, it would be interesting to analyze the implications of (\ref{eq:result2}) for gravitational corrections.

It seems strong sub-additivity does not allow us to examine the other terms --in particular we cannot get to the terms that are universal for CFTs in $d\geqslant 5$. However, this suggests that the renormalization of $\Delta \mu_{d-k}$ may have alternating signs $(-)^{k/2}$. We have shown this for $k=2, 4$, that in low dimensions give the $c$, $F$, and $a$ theorems. The statement for the last term in the expansion of the entropies of spheres  corresponds to the irreversibility of the RG in any dimensions. This sign is in agreement with the expected alternating signs of the universal coefficients. 

Let us conclude by discussing the connection with relative entropy.
The Markov property is equivalent to the cancellation 
\be
H_A+H_B-H_{A \cap B}-H_{A\cup B}=0\,
\ee
of modular Hamiltonians for a CFT \cite{Casini:2017roe}. Hence, $-\Delta S$ can be replaced by the relative entropy $S_{rel}(\rho^1||\rho^0)$ without modifying the outcome of the inequalities.  We hope to revisit these results in terms of relative entropies, extending previous work on the RG flow \cite{Casini:2016udt}. This would also include in the same scheme the $g$-theorem for CFTs with defects \cite{Casini:2016fgb}.

\section*{Acknowledgments}

We thank Marina Huerta for discussions and encouragement. 
This work was partially supported by CONICET, CNEA, and Universidad Nacional de Cuyo, Argentina. H.C. acknowledges an ``It From Qubit" grant of the Simons Foundation. 
G.T. is also supported by ANPCYT PICT grant 2015-1224.

\bibliography{EE}{}
\bibliographystyle{utphys}

\end{document}